\documentclass[twocolumn,showpacs,preprintnumbers,amsmath,amssymb,prl]{revtex4}

\usepackage{graphicx}
\usepackage{dcolumn}
\usepackage{bm}

\begin{document}

\preprint{}

\title{Quantum Conductance Steps in Solutions of Multiwalled Carbon Nanotubes}

\author{A. Urbina}
 \affiliation{Departamento de Electr\'onica, Tecnolog\'\i a de Computadoras y 
Proyectos. Universidad Polit\'ecnica de Cartagena, C/Doctor Fleming s/n, 30202 
Cartagena, Murcia. Spain}
 \email{aurbina@upct.es}
\author {I. Echeverr\'\i a}
 \affiliation{Department of Chemistry. University of Wisconsin-Madison, 1101 
University Ave., Madison, WI 53706}
\author{A. P\'erez--Garrido}
\author{A. D\'\i az--S\'anchez}
 \affiliation{Departamento de F\'\i sica Aplicada. Universidad Polit\'ecnica de 
Cartagena, C/Doctor Fleming s/n, 30202 Cartagena, Murcia. Spain}
\author{J. Abell\'an}
 \affiliation{Departamento de F\'\i sica. Universidad de Murcia, 30071 Murcia. 
Spain}

\date{\today}

\begin{abstract}
We have prepared solutions of multiwalled carbon nanotubes in Aroclor 1254, a 
mixture of polychlorinated biphenyls. The solutions are stable at room 
temperature. Transport measurements were performed using a scanning--tunneling 
probe on a sample prepared by spin--coating of the solution on gold substrates. 
Conductance steps were clearly seen. An histogram of a high number of traces 
shows maximum peaks at integer values of the conductance quantum $G_0 = 
2e^2/h$, demonstrating ballistic transport at room temperature along the carbon 
nanotube over distances longer than $1.4\mu m$.
\end{abstract}

\pacs{
72.80Rj, 
73.23Ad, 
73.63Fg. 
}

\keywords{carbon nanotubes, quantum conductance, ballistic transport}

\maketitle

Since their discovery \cite{Iijima91} theoretical calculations and experimental 
measurements suggest the enormous potential of carbon nanotubes as the building 
blocks of future molecular nanoelectronic devices \cite{Saito98, Venema99a, 
Chico96a, Perez-Garrido02}. The electronic properties of carbon nanotubes 
depend strongly on their chirality and diameter. All armchair $(n,n)$ nanotubes 
and those with $n-m$ multiple of three are metallic, the other ones are 
semiconducting with an energy gap inversely proportional to their diameter. 
This was confirmed experimentally by Wildoer and collaborators \cite{Wildoer98} 
using a scanning tunneling microscope (STM). They were able to obtain 
atomically resolved images and to perform local spectroscopy on their samples 
using the tip of the microscope as local probe. 

Electronic transport
measurements on individual single--wall carbon nanotubes (SWNT) deposited on 
metallic electrodes demonstrated their potential as molecular quantum wires 
\cite{Ebbesen96, Tans97, Tans98}. Other researchers have done experiments on 
individually manipulated multiwalled nanotubes (MWNT): Langer and collaborators 
found effects of localization and universal conductance fluctuations at $T = 
20mK$ \cite{Langer96}, and Frank et al. found quantization of the conductance 
at room temperature \cite{Frank98}. 
In the absence of scattering, the momentum relaxation length and the 
localization length are much larger than the wire length, and the transport is 
ballistic. This is the case of carbon nanotubes in the absence of lattice 
defects, where the wavefunction of the electron is extended over the nanotube, 
and there are only two channels which contributes to the electronic transport 
giving  $G=2G_0$ (where $G_0 = 2e^2/h$ is the conductance quantum) 
\cite{Saito98, Chico96a}.
If we consider elastic scattering, howewer the conductance of the system is 
described by the Landauer formula, which applies if the wavefunction can 
spread over the whole sample and is given by
\begin{equation*}
G = \frac{2e^2}{h} M\mathcal{T} = 
\frac{2e^2}{h} \sum_{\alpha \beta}^{M}|t_{\alpha \beta}|^2,
\end{equation*}
where $\mathcal{T}$ is the transmission probability for a channel to go from 
one electrode to another. This probability is given by the sum over the number 
of channels, $M$, of transmission probabilities 
from $\alpha$ to $\beta$ channel, $|t_{\alpha \beta}|^2$ (for a single--walled 
nanotube is $M = 2$). The elastic scattering affects the transmission 
probabilities and thereby reduces the conductance, which is no longer exactly 
quantized.

Molecular electronics based on the ensemble of thousands of 
nanoscale devices cannot be a real alternative to current microelectronics if 
the devices have to be individually manipulated using local probes. 
Some groups have tried a new approach using self--assembly of the 
macromolecules into controlled nets of devices or regular structures that could 
be later processed using nanolithographic methods, but carbon 
nanotubes have resisted this kind of treatment because of their difficulty to 
be diluted in appropriate solvents. Chen and collaborators studied 
both ionic and covalent solution--phase chemistry with concomitant modulation 
of the SWNT band structure \cite{Chen98}, and Liu \cite{Liu98} and Vigolo 
\cite{Vigolo00} have prepared stable colloidal suspensions in water with the 
help of surfactants allowing for collective manipulation and orientation of the 
nanotubes. Howewer, the interaction between carbon nanotubes and their organic 
solvents or between carbon nanotubes and polymers in composites or other 
nanostructured materials is not yet fully understood. 

We prepared our samples with purified arc--discharge soot material
containing carbon nanotubes, which were first submitted to a sonication bath in 
1,2--dichloroethane to unravel the material into nanotube bundles and single 
nanotubes. After evaporation of the 1,2--dichloroethane the remaining black 
powder was dissolved in a mixture of poly--chlorinated biphenyls (Aroclor 
1254), using sonication and heat treatment to improve the dilution of the 
carbon material in the viscous solvent. Aroclor 1254 is a solvent of strong 
temperature--dependent viscosity. Whith our treatment, we arrived at an 
homogeneous dark solution of carbon nanotubes in Aroclor 1254 which is highly 
stable: no precipitation has been observed for months at room temperature. 
Scanning electron microscopy showed that our nanotubes have diameters ranging 
from 15 to 25nm, and lengths ranging form 0.8 to $2 \mu m$, with a more or less 
Gaussian distribution centered at $1.4 \mu m$.
Our estimations of the critical density, $d_c$, for different solutions of 
carbon nanotubes with length $l = 1.5\mu m$, give the following values: if they 
were 
single walled (10,10) nanotubes, $d_c = 1.43\times 10^{-6}g/cm^3$, if they were 
multiwalled with ten layers, then $d_c = 1.12\times 10^{-5}g/cm^3$. 

In order to prepare our samples for the transport measurements, we used a 
solution of density $c=0.005g/cm^{-3}$ which is at least $10^3$ times bigger 
than the overlaping concentration $C^*$ of the nanotubes. The presence of 
overlaping in our samples is important to provide a percolative path between 
the nanotubes in solution and the gold substrate.
We dropped the solution mentioned before on a gold substrate and then spin 
coated it to reach an homogeneous layer of about $0.5\mu m$. We used a scanning 
probe microscope (SPM) to make contact with the nanotubes and measure the 
conductance. The platinum tip was mechanically etched and mounted on a rigid 
STM piezo. We recorded the current as a function of time and tip distance to 
the sample. The  applied bias voltage was kept fixed, so that conductance could 
be determined as a function of the position of the nanotube contact.
We start the experiment with a coarse approach of the tip to the sample. Once 
electrical contact has been established, the tip is piezo controlled by the SPM 
and driven in and out cyclically. The vertical travel used is $2\mu m$. Data 
from sequences of 50 to 100 cycles were 
recorded automatically. When the tip makes contact with the sample, one 
nanotube can stick to it, remaining stuck while the tip cycles, and thus 
providing an electrical contact through the nanotube with the gold substrate, 
directly or through a percolative path with other nanotubes of the solution. 
While the tip is cycling, the contact of the nanotube with the tip remains 
constant, otherwise the nanotube contact with the substrate is changing while 
it is being lifted by the tip. When the tip retracts a distance larger than the 
length of the nanotube, the contact is broken and the conductance goes to zero. 
We can record sequences of more than fifty cycles, but usually it is difficult 
to reach more than seventy cycles without loosing contact with the nanotube.

\begin{figure}
\includegraphics[width = \columnwidth]{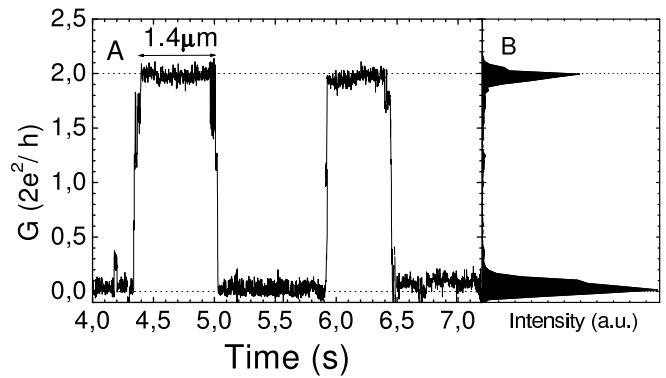}
\caption{\label{fig1} A sequence of measurements of conductance versus time 
showing steps at $2G_0$, where $G_0$ is the conductance quantum $G=2e^2/h$. Two 
steps of different length out of a sequence of more than fifty are shown. The 
histogram shows the persistence of the quantized value along the sequence, with 
a clearly defined peak at $2G_0$. The bias voltage between tip and substrate 
was $V_p =10mV$.}
\end{figure}

In figure \ref{fig1} we plot the conductance $G$ versus time for a portion of a 
sequence in which two steps of different length are observed. This is 
representative of a high number of steps, which can be used to construct an 
histogram for every sequence, shown in figure \ref{fig1}B. The histogram is 
taken over the first fifty cycles. In this figure, a 
peak centered at the conductance value of $2G_0$ can be seen, along with the 
broadening expected from the noise in our measurement.
Note that if the nanotubes behaved as diffusive conductors, we would expect the 
conductivity to vary as a function of the distance between the nanotube contact 
points to the substrate and the tip. Howewer, what we 
observe in our experiment is a series of plateaus, meaning that the 
conductivity remains constant for periods of time, which in our experiment are 
equivalent to tip displacements ranging from 0.8 to 1.4$\mu m$. This can be 
explained by changes in the contact point between nanotube and substrate, which 
sometimes might even involve mechanical deformation (bending) of the nanotube. 
The value of the conductance remains fixed while the current path is changing, 
thus the nanotube behaves like a ballistic conductor. Furthermore, the 
conductance appears to be quantized, taking values only at even integer 
multiples of the 
conductance quantum $G_0$. This result suggests that the nanotube behaves like 
a quantum wire, with only two conductance channels. We find this to be true, 
regardless of the diameter and length of the nanotubes involved. We attribute 
the different plateau lengths observed to differences in the length of nanotube 
between tip and substrate, which could be due to the adherence of a 
different nanotube or to a occasional displacement of the previously stuck one 
along the platinum tip. The presence of the solvent may enhance nanotube 
adherence to the tip. Howewer, as the contact heats up, the viscosity of the 
Aroclor decreases modifying the nanotube--tip structure in an unpredictable 
way. Nevertheless, the value of the conductance remains fixed at $2G_0$. The 
persistence of this value indicates that only the outer layer contributes to 
the electronic transport. The inner layers are isolated from the outer because 
the resistivity in the direction perpendicular to the tube axis is very high, 
as one could expect from similarities with graphite.

\begin{figure}
\includegraphics[width = \columnwidth]{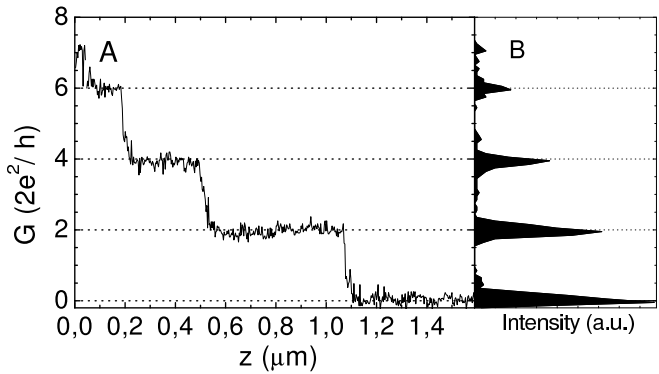}
\caption{\label{fig2} In some of the sequences of conductance, three 
steps appeared at even integer values of the conductance quantum $G_0$. The 
conductance is plotted versus distance between tip and contact point with the 
sample. Two peaks are clearly seen in the histogram. A third peak is blurred by 
the noise. The bias voltage between tip and substrate was $V_p =25mV$.}
\end{figure}

A second kind of results obtained with the same experimental procedure can be 
seen in figure \ref{fig2}. A series of steps at even integer values 
of the conductance quantum $G_0$ appears when the current along the nanotube is 
plotted versus the position of the tip. Since in single--walled metallic 
nanotubes only two channels contribute to electronic transport, we think that 
the new steps come from transport through other nanotubes. It is possible that 
two or more nanotubes get stuck to the tip, thus providing more than one 
connection, also ballistic and quantized, for the current between tip and 
substrate to flow. When the tip makes contact, sometimes the staircase begins 
with a step at a value of 4 and occasionally 6 $G_0$. When the tip begins to 
retrieve, the step remains fixed for a while and suddenly changes reducing to 
the next even integer value of $G_0$. We have never observed a step which 
changes in values higher than $2G_0$. This provides support to the idea that 
only the outer shell of the MWNT contributes, because if two (or more) 
consecutive metallic shells were contributing, it would provide a step of 
$4G_0$ (or higher) when this nanotube losses contact with the gold substrate. 
An histogram of twenty cycles is shown in figure \ref{fig2}B, showing peaks at 
even integer values.

Some of the observed steps present some slight but clear deviations from the 
quantized values, even after considering the noise of our measurements. This 
effect is more pronounced at the end of the step, when the current flows 
through longer distances along the nanotube. We attribute this deviation to the 
elastic scattering produced by structural defects of the nanotube atomic 
lattice, which affects the transmission coefficients and deviates the 
conductance from the quantized value as described by the Landauer formula. Our 
nanotubes were subjected to ultrasonic bath and this procedure could have 
produced some structural damage in the nanotubes. Such specific defects may 
affect the conduction at special energies werheas the transport remains 
unaffected for different energies.

An alternative explanation of the elastic scattering could be found in the 
interaction between shells of the multi--walled nanotube. The shell which is 
contributing to the electronic transport can be affected by its interaction 
with other shells, introducing an effective potential equivalent to structural 
disorder. Since we can not resolve the chirality of the successive shells of 
the MWNT, we could not study this effect. We think that the interaction between 
shells can be in part responsible for the discrepancy between our results and 
those found by S.~Frank and coworkers (Ref. \cite{Frank98}). They found a 
quantized conductance at one unit of the conductance quantum $G_{0}$ instead of 
the expected $2G_{0}$. They attached a carbon fiber with protruding carbon 
nanotubes to a gold wire and used it as the tip of a scanning probe 
microscope. This fiber was found in the soft material inside the hardshelled 
deposit of an arc-discharge and no aditional treatment was performed to the 
fiber. We think that different shells of their MWNT can have electrical contact 
with the gold wire, increasing the intershell interaction even if only the 
outer shell have contact with the mercury melt. Recent theoretical works 
\cite{Roche01a, Roche01b} have studied the effect of intershell coupling and 
incommensurability in MWNT and they found that the mixing of quantum channels 
results in an enhanced contribution of backscattering and quantum interference 
effects which gives a quantum correction that reduces the conductivity. 
This effect could explain the different kinds of electronic transport reported 
for MWNT: ballistic, diffusive and even insulating behaviours. This may be the 
case of Ref. \cite{Frank98}, but the exact reduction of one unit of the 
conductance quantum remains unexplained, unless the mechanism mentioned above 
eliminates only one of the two channels in the outer shell of the MWNT.

We also measured the current--voltage characteristics for some samples. When an 
electrical contact was obtained, we varied the bias voltage in a range between 
0.1 and 1V. The I/V plot showed a linear behavior, which indicates that the 
nanotubes are indeed metallic. Howewer, at higher voltages ($V \geq 0.6V$) we 
observed some instabilities even if there was no sign of current saturation, 
those inestabilities may arise from the contribution of different subbands even 
if a single shell is contributing to transport. For a bias 
voltage of 1V, the current density is on the order of $J\sim 10^6 Acm^{-2}$, 
depending on the diameter of the nanotube. The dissipated power is $\sim 1mW$. 
The electrical breakdown of the nanotube occurs at higher values of bias 
voltage, when the non linear regime is reached. It has been reported that for 
MWNT this breakdown occurs at powers of $300 \mu W$ \cite{Collins01b}. The 
onset of saturation and the eventual breakdown are linked to a common 
dissipative process involving the excitation of phonons, temperature raising in 
the nanotube and a process of oxidation and loss of successive carbon shells. A 
simple heat transfer analysis, taking the nanotube as a solid cylinder with a  
bulk thermal conductivity of $\kappa = 3000Wm^{-1}K^{-1}$\cite{Kim01} gives for 
a nanotube of $1.4\mu m$ long and 20nm in diameter a temperature $T = 1850K$, 
which is not possible, because nanotubes start to oxidate at $T\sim 1000K$. If 
we take other reported values such as $\kappa = 25Wm^{-1}K^{-1}$ \cite{Yi99}, 
which possibly apply for nanotubes with more defects, or higher diameters, as 
is our case, the temperature would reach $T\sim 22000K$. Clearly heat is not 
dissipated in the nanotube, further supporting that the transport is ballistic.

In summary, we have prepared stable solutions of multiwalled carbon 
nanotubes in Aroclor 1254. Transport measurements using a scanning 
probe showed ballistic transport through distances longer than $1.4 \mu m$ and 
conductance steps at even integer values of the conductance quantum $G_0$.

\begin{acknowledgments}
We wish to acknowledge Professor John Schrag for supplying Aroclor 1254, and 
for helpful discussions and suggestions. This work has been partially supported 
by Fundaci\'on S\'eneca under Grant No.PI-60-00858-FS-01.

\end{acknowledgments}

\bibliography{/home/urbina/fisica1/CNTproposal/biblioCNT}

\end{document}